\documentstyle[twocolumn,epsf,11pt]{article}

\begin{document}

\onecolumn

\begin{titlepage}

\setcounter{page}{0}

\thispagestyle{empty}

\vfill

\begin{flushright}
\vspace*{15mm}
{\tt hep-lat/9305010}
{\tt RIKEN-AF-NP-154}
\end{flushright}

\vfill

\begin{center}
{\Large Phases of the three-state Potts model in three spatial dimensions}
\\[15mm]
Shigemi Ohta\\
The Institute of Physical and Chemical Research (RIKEN)\\
Wako-shi, Saitama 351-01, Japan

\end{center}

\vfill

\begin{abstract}
The three-state Potts model is numerically investigated on
three-dimensional simple cubic lattices of up to \(128^3\) volume,
concentrating on the neighborhood of the first-order phase transition
separating the ordered and disordered phases.  In both phases clusters
of like spins are observed with irregular boundaries.  In the ordered
phase the two different non-favored spins are found to attract each
other with a long but finite range.  As a result, the neighborhoods of
the non-favored spins are interpreted as domains of the disordered
phase.  This explains why the first-order phase transitions associated
with the global \(Z_3\) symmetry, including the SU(3) pure-gauge one,
are so weak.
\end{abstract}

\vfill

\end{titlepage}

\twocolumn

Finite-temperature systems with global \(Z_3\) symmetry in three
spatial dimensions have in common a weak first-order phase transition
separating ordered and disordered phases.  Lattice numerical
calculations of the \(SU(3)\) pure-gauge quantum chromodynamics (QCD)
at finite temperature \cite{ref:puregauge} revealed that a first-order
phase transition separates a low-temperature color-confining phase and
a high-temperature non-confining phase.  The confining phase is
disordered with the Polyakov line order parameter taking zero
expectation value.  The non-confining phase is ordered and the
Polyakov line takes a finite expectation value falling on one of the
three \(Z_3\) axes in the complex plane.  The transition is considered
weak because the latent heat is small and the Polyakov line
correlation length grows.  Numerical simulations of the three-state
Potts model \cite{ref:Potts} showed there is a first-order phase
transition separating a low-temperature ordered phase and a
high-temperature disordered one.  This transition is weak in the sense
the latent heat is small.  However, it is not yet understood why these
phase transitions are so weak.  This letter reports the author's
investigation of this problem.

We consider the simplest form of the three-state Potts model on
three-dimensional simple cubic lattices.  At each site, \(x\), of the
lattice, a spin, \(\sigma_x\), is defined.  It takes one of the three
possible states in the \(Z_3\) group,
\(\sigma_x \in \{0, 1, 2\}\).  The thermodynamics of the system is
described by the partition function,
\[
Z = \exp(-H/T) = \exp(J \sum_{\langle xy \rangle} \delta(\sigma_x,
\sigma_y)),
\]
where the sum runs the nearest-neighbor pairs \(\langle xy \rangle\)
and Kronecker's symbol is defined as \(\delta(\sigma_x, \sigma_y) =
1\) if \(\sigma_x = \sigma_y\) and 0 otherwise.  The model is
invariant under global \(Z_3\) transformations.  Here we consider only
the ``ferromagnetic'' couplings, \(J > 0\).  The effective
interactions among the Polyakov lines in the finite-temperature
pure-gauge QCD are of this type \cite{ref:FOU}.

\begin{figure}[tb]
\begin{center}
\leavevmode
\epsfxsize=75mm
\epsffile{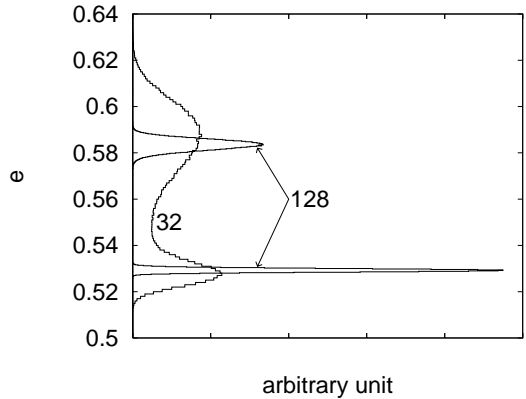}
\caption{Distribution of the dimensionless internal energy density
from heatbath simulations at \protect\(J = 0.5505\protect\) on
\protect\(32^3\protect\) (32) and \protect\(128^3\protect\) (128)
volumes.}
\label{fig:bothhist}
\end{center}
\end{figure}

Earlier numerical simulations showed that a weak first-order phase
transition that separates a low-temperature ordered phase and a
high-temperature disordered phase at \(J \simeq 0.5505\).  The phase
transition is considered weak because the dimensionless internal
energy density, \(e \equiv \langle \delta (\sigma_x, \sigma_y)
\rangle\), shows only a small gap at the phase transition, from about
0.58 to 0.53, compared with the maximum possible gap from 1 to 1/3.
See Fig.\ \ref{fig:bothhist} for distribution of this quantity.  These
results were, however, obtained on relatively small volumes such as
\(32^3\).  The two broad peaks marked as ``32'' in Fig.\
\ref{fig:bothhist} shows the distribution from such a small-volume
simulation reproduced by the author.  Although the two-peak structure
is clear enough for establishing the first-order nature of this phase
transition, the significant overlapping of the broad tails in the
middle reflects the fact that flip-flop transitions between the two
phases occur too frequently and it is hard to tell in which of the two
phases the system resides at a given instant.  Because of this
difficulty, the natures of the individual phases could not be studied,
nor the reason why the phase transition is so weak.

To avoid this difficulty one has to simulate the system on larger
volumes.  The author wrote a heatbath simulation code for the model
that runs on an experimental parallel computer, AP1000, built by
Fujitsu Laboratory \cite{ref:AP1000}.  On the largest existing
configuration of the computer with \(16 \times 32 = 512\)
microcomputer ``cells,'' the code performs one million heatbath
updates of a \(128^3\) volume in about 32 hours.  Typically a couple
of million heatbath updates are made at four values of the coupling,
\(J = 0.55025\), 0.5505, 0.55075, and 0.551.  At each of these
coupling values, two independent simulations were made; one starting
from a completely ordered configuration and the other starting from a
completely disordered one.

At the highest temperature, \(J = 0.55025\), the ordered-start
simulation quickly converged to the disordered phase within 100K
heatbath updates, while the disordered-start simulation remained in
the disordered phase indefinitely through the duration of the
simulation for more than one million updates.  In contrast, at the
lowest temperature, \(J = 0.551\), the disordered-start simulation
quickly converged to the ordered phase while the ordered-start one
remained in the ordered phase indefinitely.  Coexistence of the two
phases is observed at both of the intermediate temperatures, \(J =
0.5505\) and 0.55075: The ordered-start simulations remained in the
ordered phase and the disordered-start ones remained in the disordered
phase through their durations of more than two million heatbath
updates.  The corresponding distribution of the internal energy
density is shown in Fig.\ \ref{fig:bothhist}, marked as ``128.''  The
two peaks are now completely separated, and there is no doubt about in
which of the phases the system resides: We are now able to investigate
the natures of the individual phases.

The two-point correlation function, \(C_{ij} (r)\), is defined as the
probability to find a pair of spin \(i\) and \(j\) separated by the
distance \(r\).  They are calculated once in every 100K heatbath
updates, and are proven free of autocorrelation at least in the
relevant range of \(r \le 8\).

\begin{figure}[tb]
\begin{center}
\leavevmode
\epsfxsize=75mm
\epsffile{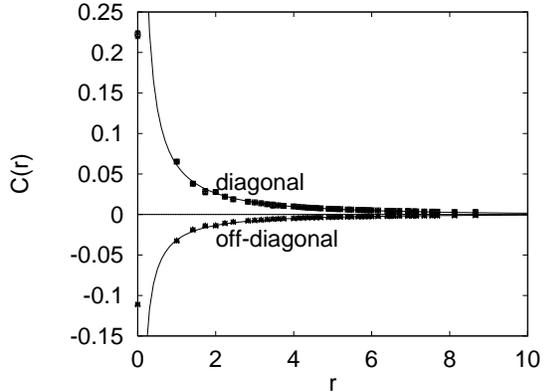}
\caption{Correlations in the disordered phase at \protect\(J =
0.5505\protect\).  The constant terms \protect\(p_i p_j\protect\) are
subtracted.  The curves are obtained by the
least-\protect\(\chi^2\protect\) fittings to the Yukawa form
summarized in Table \protect\ref{tab:J5505d}.}
\label{fig:J5505d_corr}
\end{center}
\end{figure}

In the disordered phase the correlations are expected
to decay exponentially in Yukawa form 
\[
C_{ij} (r) \rightarrow \alpha_{ij} \frac{\exp(-m_{ij}r)}{r} + p_i p_j,
\]
as the distance \(r\) increases toward infinity \cite{ref:Hintermann}.
The constant term \(p_i p_j\) is given by the product of
probabilities, \(p_i\) and \(p_j\), to find a site with spin \(i\) and
\(j\) respectively.  In the disordered phase they should all approach
1/3 as the system size increases, because the \(Z_3\) symmetry is
preserved.  For the same reason, there are only two different sets of
correlations: the diagonal ones, \(C_{00} = C_{11} = C_{22}\), and the
off-diagonal ones, \(C_{01} = C_{02} = C_{12}\).  Moreover, because of
the conservation of probability,
\begin{equation}
C_{i0} (r) + C_{i1} (r) + C_{i2} (r) = 1, \label{eqn:sumrule}
\end{equation}
for any \(i\), all the correlation mass must agree with each other and
the amplitudes obey a simple relation \(\alpha_{00} = \alpha_{11} =
\alpha_{22} = -2 \alpha_{01} = -2 \alpha_{02} = -2 \alpha_{12}\).  We
numerically confirmed that all the correlations are fitted well by the
Yukawa form, as can be seen in Fig.\ \ref{fig:J5505d_corr} where the
correlations are shown with the constant terms subtracted.  Within
statistical errors, the fitting parameters (see Table
\ref{tab:J5505d}) satisfy the above requirements.  The positive
\begin{table}[b]
\begin{center}
\caption{Two-point correlations in the disordered phase at \protect\(J
= 0.5505\protect\).  Two-parameter least-\protect\(\chi^2\protect\)
fit to the Yukawa form for the range \protect\(r \ge 3\protect\).}
\label{tab:J5505d}
\begin{tabular}{lrr}
\hline\hline
\protect\((ij)\protect\)
& \multicolumn{1}{l}{\protect\(m\protect\)} &
\multicolumn{1}{l}{\protect\(\alpha\protect\)} \\
\hline
(00), (11), (22) & 0.15(1) &  0.071(3) \\
(01), (02), (12) & 0.16(1) & -0.037(1) \\
\hline\hline
\end{tabular}
\end{center}
\end{table}
diagonal amplitudes, \(\alpha_{ii} > 0\), shows the like spins attract
each other, in accordance with the ferromagnetic interaction.  The
corresponding correlation mass of \(m_{ii} \simeq 0.15\) suggests
there exist clusters of like spins of the size of several lattice
spacings.  Indeed in the spin distributions one sees many such
clusters.  It is interesting to note that these clusters are neither
smooth nor convex in shape, but are complex and concave.

\begin{figure}[t]
\begin{center}
\leavevmode
\epsfxsize=75mm
\epsffile{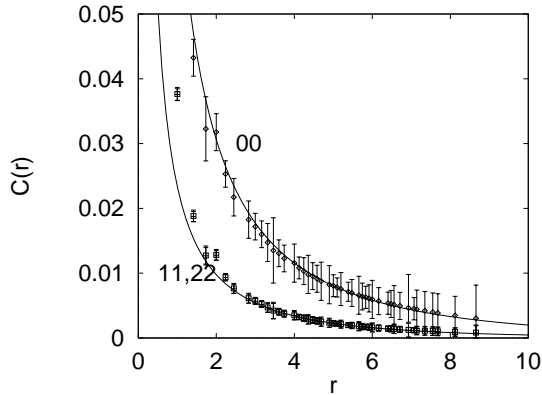}
\caption{Diagonal correlations in the ordered phase at \protect\(J =
0.5505\protect\).  0 means the favored spin and 1 and 2 are
non-favored.  The constant terms are subtracted.  The curves
are from least-\protect\(\chi^2\protect\) fitting summarized in Table
\protect\ref{tab:J5505o}.}
\label{fig:J5505o_dia}
\end{center}
\end{figure}

\begin{figure}[t]
\begin{center}
\leavevmode
\epsfxsize=75mm
\epsffile{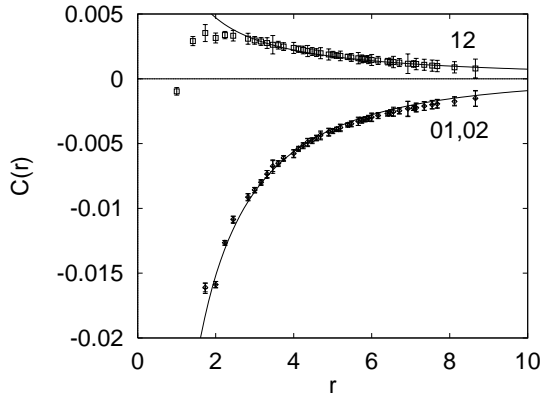}
\caption{Off-diagonal correlations in the ordered phase at \protect\(J
= 0.5505\protect\).  The constant terms are subtracted.  The curves
are from least-\protect\(\chi^2\protect\) fitting summarized in Table
\protect\ref{tab:J5505o}.}
\label{fig:J5505o_off}
\end{center}
\end{figure}

In the ordered phase there is one favored spin which we label as 0,
and two non-favored ones which we label as 1 and 2.  No analytic
result is known about the behavior of the correlations.  The current
numerical results show the Yukawa form fits them well also in this
case.  See Fig.\ \ref{fig:J5505o_dia} and \ref{fig:J5505o_off}, where
the constant terms are subtracted again.  The fitting parameters are
given in Table
\ref{tab:J5505o}.
\begin{table}[b]
\begin{center}
\caption{Two-point correlations in the ordered phase at \protect\(J
= 0.5505\protect\).  Two-parameter least-\protect\(\chi^2\protect\)
fit to the Yukawa form for the range \protect\(r \ge 3\protect\).  The
favored spin is 0.}
\label{tab:J5505o}
\begin{tabular}{lrr}
\hline\hline
\protect\((ij)\protect\)
& \multicolumn{1}{l}{\protect\(m\protect\)} &
\multicolumn{1}{l}{\protect\(\alpha\protect\)} \\
\hline
(00)       & 0.14(2) &  0.081(3) \\
(11), (22) & 0.18(3) &  0.028(4) \\
(01), (02) & 0.15(1) & -0.041(1) \\
(12)       & 0.03(3) &  0.010(1) \\
\hline\hline
\end{tabular}
\end{center}
\end{table}
The positive amplitudes of the diagonal correlations, \(\alpha_{ii} >
0\), are in accordance with the ferromagnetic interaction.  Smaller
correlation mass for the favored spin, 0, compared with those of
non-favored spins, 1 and 2, may not be statistically significant, but
nonetheless is consistent with the fact that the favored spin
dominates the volume in the ordered phase.  At this coupling, \(J =
0.5505\), about 60 \% of the volume is covered by the favored spin,
while the remaining is evenly split between the two non-favored ones.
On the other hand the negative amplitudes of the correlations between
the favored spin and either of the two non-favored spins,
\(\alpha_{01} < 0\) and \(\alpha_{02} < 0\), means the favored spin
repel the non-favored spins.  These suggest that there form clusters
of like spins, which are indeed observed in the spin distributions.
Again the clusters have irregular boundaries \cite{ref:ohta92}.

The correlation between the two different non-favored spins, \(C_{12}
(r)\), is the most interesting.  It starts from zero at the origin as
it should be, but then overshoots the asymptotic value \(p_1 p_2\),
and approaches it from above (see Fig. \ref{fig:J5505o_off}.)  This
means an attractive force acts between the two different non-favored
spins.  The attractive part alone can be fitted by the Yukawa form if
we limit the range of fitting to \(r \ge 3\) (Fig.\
\ref{fig:J5505o_off}.)  Note that the correlation mass, \(m_{12} =
0.03 \pm 0.03\), is much smaller than those for the other cases, the
smallest of which is \(m_{00} = 0.14 \pm 0.02\).  It is not
appropriate, however, to consider this as a truly long-range
correlation with zero correlation mass.  Again because of the
conservation of probability (\ref{eqn:sumrule}) the correlation mass
\(m_{12}\) in the long-distance limit \(r \rightarrow \infty\) must be
equal to the smaller of the masses \(m_{01}\) and \(m_{11}\).  Much
smaller value of \(m_{12}\) we are seeing must come from subtle
balance of the two terms \(C_{01} (r)\) and \(C_{11} (r)\) in the
range we are looking at, and has to give way to the true correlation
mass at some longer range.  For this reason, we will call this
attractive part as the ``middle-range'' attraction in the
remainder of this letter.

Yet it is useful to explore what would happen if we had a really
long-range attraction between the two different non-favored spins:
such an attraction is likely to cause instability of the ordered
phase, because the repulsion against the non-favored spins from the
favored ones can then be compensated by paring the two different
non-favored spins attracting each other.  Hence we expect that the
smaller the correlation mass \(m_{12}\), the weaker the phase
transition.  Indeed in the spin distribution of this phase one finds
many clusters of the non-favored spins.  And such a non-favored
cluster almost always accompany clusters of the other non-favored spin
in its neighborhood.  It should be noted that such pairing of the
clusters of the two different non-favored spins is a way to maintain
the global \(Z_3\) symmetry which requires the volumes occupied by the
two non-favored spins must be equal.

\begin{figure}[tb]
\begin{center}
\leavevmode
\epsfxsize=75mm
\epsffile{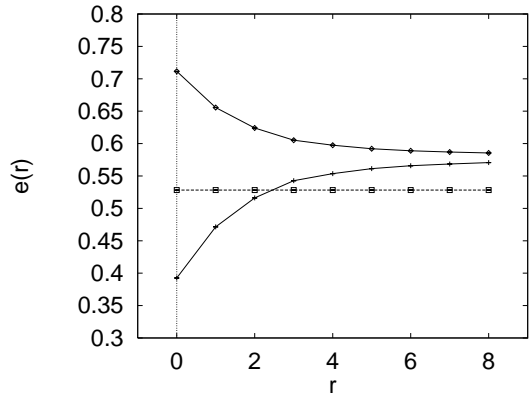}
\caption{Local-averaged internal energy density at  \protect\(J =
0.5505\protect\).  Solid lines are from the ordered phase, around the
favored spin (upper one) and non-favored spin (lower one).  Dashed line
is from the disordered phase.  The errors are smaller than the symbol
sizes.}
\label{fig:J5505_ie}
\end{center}
\end{figure}

As was mentioned earlier, these clusters of the non-favored spins
appear in irregular shapes.  The neighborhood of such a cluster of
complex-shaped clusters of the non-favored spins can be interpreted as
an island of the disordered phase in the sea of the ordered phase.
This is clearly seen by looking at the local-averaged internal energy
density, \(e_i (r)\), which is defined in the sphere of radius \(r\)
with spin \(i\) at its center.  See Fig.\ \ref{fig:J5505_ie} for its
dependence on the radius.  In the disordered phase, the quantity does
not show any dependence and stays at its asymptotic value of \(\simeq
0.53\).  In the ordered phase, the one around the favored spin
approaches the asymptotic value of \(\simeq 0.58\) monotonously from
above, while the other around the two non-favored spins starts from a
value smaller than in the disordered phase and approaches the
asymptotic value monotonously from below.  These disordered domains
increases the internal energy density of the ordered phase and
explains why the latent heat is so small.

Irregular boundaries of the non-favored clusters and disordered
domains in the ordered phase near the transition may be consistent
with very small surface tension at the confined-deconfined phase
boundaries found by a MIT-bag calculation \cite{ref:Mardor} as well as
by lattice QCD numerical calculations \cite{ref:BandH}.  Note,
however, that the irregular boundaries themselves suggests inadequacy
of the sharp spherical boundaries assumed or imposed in these
calculations.

The ``middle-range'' attractive correlation between the two
different non-favored spins remains attractive down to the lowest
temperature simulated, \(J = 0.551\).  Its range, however, decreases
to \(m_{12} = 0.07(3)\) at \(J = 0.55075\) and 0.11(3) at 0.551.  This
way the ordered phase away from the first-order transition
consolidates itself.

It would be interesting to check if the above mechanism of weakening
the first-order phase transition works in the pure-gauge QCD
thermodynamics.  As a first step, the author generated several hundred
pure-gauge configurations on a \(16^3 \times 2\) lattice by the
``APE6'' computers in Rome, using a hybrid Monte Carlo algorithm
written by the APE group \cite{ref:Donini}.  From each gauge
configuration a distribution of Polyakov lines is calculated.  Near
the first-order phase transition, most of the Poliyakov lines are
found in the neighborhood of the three \(Z_3\) axes with non-zero
magnitude, even in the disordered phase.  Each Polyakov line is
projected onto the nearest of the \(Z_3\) axes, resulting in
distributions of the Potts-model spins.  Correlations of the spins are
then calculated, and found to behave in the same manner as their
counterparts in the three-state Potts model.  Most notably, the
correlation between the two different non-favored spins in the ordered
phase shows the same kind of ``middle-range'' attraction.  This
allows one to speculate that the deconfined phase of QCD near the
phase transition is susceptible to mixture of complex-shaped
``droplets'' of the confined phase \cite{ref:Mardor}.  If such is
indeed the case, perturbation calculations of the QCD deconfined phase
would be useless near the phase transition.  Hence it is important to
investigate this point in more detail.

The author thanks Parallel Computing Research Facility, Fujitsu
Laboratory, for computing time on their AP1000 parallel computer on
which much of this work was done.  He is also thankful for fruitful
discussions with Norman Christ, Enzo Marinari, Giorgio Parisi, and
Roberto Petronzio.  Thanks are also due to the hospitality extended to
him by the APE100 group, especially by Federico Rapuano and Raffaele
Tripiccione.  Finally, he warmly thanks Andrea Donini and Stefano
Antonelli for kindly allowing him to use their hybrid Monte Carlo
simulation code for the APE computer.

\end{document}